\documentclass[12pt,aps,floats]{revtex4}
\usepackage{graphicx}
\usepackage{epsfig}
\usepackage{axodraw}

\begin{document}
\title{Probing the Planck Scale with Neutrino Oscillations}

\author{Ram Brustein, David Eichler, Stefano Foffa}
\address{Department of Physics,
Ben-Gurion University, Beer-Sheva 84105, Israel\\ email:
ramyb,eichler,foffa@bgumail.bgu.ac.il}


\begin{abstract}

Quantum gravity ``foam", among its various generic Lorentz
non-invariant effects, would cause neutrino mixing.  It is shown here
that, if the foam is manifested as a nonrenormalizable effect at scale
$M$, the oscillation length generically decreases with energy $E$ as
$(E/M)^{-2}$.  Neutrino observatories and long-baseline experiments
should have therefore already observed foam-induced oscillations, even
if $M$ is as high as the Planck energy scale.  The null results, which
can be further strengthened by better analysis of current data and
future experiments, can be taken as experimental evidence that Lorentz
invariance is fully preserved at the Planck scale, as is the case in
critical string theory.

\end{abstract}

\maketitle

The noisy vacuum of a quantum theory of gravity could  {\t a priori} be
imagined to have a variety of effects on the wavefunction of a particle
traveling through it. Although unitarity would probably constrain the
effects of the vacuum on the wavefunction's amplitude, one could
imagine that its phase might be shifted by the local effects of quantum
black holes and the like which quickly pop in and out of the vacuum.
Also, the type (flavor) of the particle might be affected: for example,
a virtual or quantum black hole could as Hawking has suggested
\cite{hawking}, erase information by swallowing a neutrino of one type
and spit out one of a different type as it evaporates. Thus, possible
effects that could be considered include Lorentz invariance violating
(LIV) indices of refraction \cite{AC,Bi,ACP}, flavor oscillations of
neutral particles\cite{Liu,ColGl1,ColGl2,eichler,DesGas}, and
corrugations of wavefronts, i.e. Rayleigh scattering \cite{eichler}.
Rayleigh scattering, which would
imply momentum non-conservation, appears to be ruled out
by the observation that 10 MeV neutrinos propagated without Rayleigh
scattering from supernova 1987A \cite{eichler}
 (assuming that their scattering
cross-section off cells of Planck foam scales as $E^2$ ).  We conclude
that if such ``foam" effects exist, they must preserve translation
invariance on the Planck distance scale and on larger ones;  thus any
non-vanishing effects must be cooperative and coherent over space-like
distances in this range of scales; it may cause wave dispersion but not
opalescence or translucence.

There has been considerable work on dispersive, LIV vacua. Colladay and
Kostolecky have exhaustively classified renormalizable LIV effects
\cite{kost1,kost2}. Coleman and Glashow \cite{ColGl1,ColGl2} discuss
neutrino oscillation from renormalizable LIV terms, which yield an
oscillation length that scales as $1/E$, and their results imply that
any such effect that is of order unity at the Planck scale is already
ruled out by many orders of magnitude.  Amelino-Camelia et al.
considered a photon velocity that has a linear term proportional to
$E/M$ and showed that it is marginally consistent for astrophysical
gamma ray pulses if $M$ is of order $10^{16}$ GeV.

The leading quantum gravity theory - critical string theory - predicts
that in flat (empty) space all interactions are Lorentz invariant (LI)
down to, and including, the Planck scale. LI in flat space is crucial
to the internal mathematical consistency of string theory in that it
guarantees that its symmetries are valid at the quantum level (see, for
example, \cite{Polchinski}). However, the Universe, possessing a
preferred frame of reference, is clearly not LI. String theory allows
such spontaneous breaking of LI, by allowing the possibility that the
curvature and additional moduli fields have time- and space-dependent
expectation values, but it does not allow LIV Planck scale ``foam".

In this paper we consider a particular class of LIV effects -
{\it nonrenormalizable} effects of the vacuum, which could be induced
by a high energy/short distance physical cutoff - and show that
they give rise to a neutrino dispersion that is linear in E/M.
This is the dispersion relation considered by Amelino-Camelia et
al.~\cite{AC} for photons, but, like Coleman and
Glashow\cite{ColGl1,ColGl2}, we consider neutrino oscillations
that it could induce. In contrast to the latter, we consider
nonrenormalizable corrections, which depend on a higher power of
$E/M$. If the difference between the propagation velocities of
different neutrino flavors is proportional to $E/M$ then the
oscillation length is proportional to $(E/M)^{-2}$. (Amusingly,
for $M\sim M_{\rm Planck}$ it would give an oscillation length of
order an astronomical unit for solar neutrino energies and
atmospheric length scales for atmospheric neutrino
energies\cite{eichler}, but see below.)  We then apply the
argument\cite{eichler} that high energy experiments of neutrino
oscillations with $E^{-2}$ mixing lengths should be able to
``detect" nonrenormalizable cutoffs as high as the Planck scale.
We argue that such experiments can, by not detecting generic
effects of such cutoffs, provide experimental support for
critical string theory versus many proposed alternative theories,
as the latter imply LIV physical cutoffs, while critical string
theory provides a LI cutoff.

It is clear that our arguments are generic, and restrict or rule out
only those models of quantum foam in which neutrino flavor is
unprotected by a symmetry and Lorentz
invariance is broken by nonrenormalizable interaction terms,
as pointed out in \cite{ellis,gac}. It remains to be seen whether
any specific models of quantum gravity effects can evade our conclusions, 
as suggested in \cite{ellis,gac}.

To understand the appearance of nonrenormalizable LIV terms we
model any type of short distance interaction which may be induced
by quantum gravity foam, black holes swallowing particles and
spitting them out, ultraviolet LIV cutoff physics, etc.
First, we assume that such interaction is strong only at some
characteristic energy scale $M$ (e.g. $M_{\rm Planck}$) so it can be
modeled by the following interaction term in the effective
Lagrangian,
\begin{equation}
 \label{livint}
 {\cal L}_{\rm int}=M \int d^4 x' \psi^{\dagger\rho}_A(x)
 f^{AB}_{\rho\sigma}\left[M (x-x')\right]  \psi^{\sigma}_B(x').
\end{equation}
Here $f$ determines the interaction strength, and is therefore
assumed to be small when its argument is larger than unity,
$A,B=1,2$ label different neutrino eignestates of the total
Hamiltonian, $\rho,\sigma= 1,...,4$ are fermionic indices. To
avoid Rayleigh scattering as discussed previously we have assumed
that $f$ is a function only of the distance four-vector $x-x'$.
We may now expand $\psi^{\sigma}_B(x')= \psi^{\sigma}_B(x)+
\partial_\mu\psi^{\sigma}_B(x) (x'-x)^\mu +\frac{1}{2}
\partial_\mu\partial_\nu \psi^{\sigma}_B(x)(x'-x)^\mu (x'-x)^\nu + ...$,
where the dots stand for higher derivative terms. This expansion
may be used to convert the interaction Lagrangian (\ref{livint})
into a series of local terms. Each additional derivative comes
with an additional power of $1/M$, demonstrating that the terms
induced by (\ref{livint}) indeed depend on powers of $E/M$. The
lowest dimension nonrenormalizable terms are dimension five
operators and therefore depend on $E^2/M$. They are given by the second
moments of $f^{AB}_{\rho\sigma}$. It may well be that the
renormalizable terms which could have been induced by (\ref{livint}) are
absent. For example, if  $f^{AB}_{\rho\sigma}$ depends only  on the
absolute value  $|x-x'|$, no dimension four operators are induced. If,
in addition, neutrino masses  are protected, say, by chiral symmetry,
and $f^{AB}_{\rho\sigma}(|x-x'|)$ factors into  constant matrix
$M_{\rho \sigma}^{AB}$ and a universal kernel $f(x-x')$, then neither
are  dimension three operators induced.

Consider the diagram in Fig.~1. The effective interaction vertex
connecting the incoming and outgoing states includes, by
assumption, quantum gravity effects. Because any flavor
gravitates, the effective vertex does not generically respect the
global flavor symmetry (e.g. a quantum black hole swallowing one
type of neutrino and evaporating into another type) so in its
presence the eigenstates of the weak interactions
$|\alpha\rangle$, $|\beta\rangle$ are mixtures of the eigenstates
of total Hamiltonian $|A\rangle$, $|B\rangle$,
$|\alpha\rangle=\cos\theta |A\rangle+\sin\theta |B\rangle$, and
$|\beta\rangle=- \sin\theta |A\rangle+\cos\theta |B\rangle$, and
therefore the interaction terms obtained from (\ref{livint})
induce mixing between flavor states. We will argue shortly that
the induced mixing angles are generically large
$\sin^2(2\theta)\sim1$. (In \cite{ellis}  it is claimed that in
certain ``kinematical models" flavor mixing does not arise).

\begin{figure}
\begin{center}
\begin{picture}(75,80)(0,0)
 \ArrowLine(-30,50)(20,60) \Text(-30,60)[l]{$B,\sigma$}
 \Text(7,65)[1]{$x$}
 \Vertex(20,60){10} \Text(18,78)[l]{$f^{AB}_{\rho\sigma}$}
 \ArrowLine(25,60)(75,50)
 \Text(35,66)[1]{$x'$}
 \Text(60,60)[l]{$A,\rho$}
\end{picture}
\vspace{-.5in} \caption{\label{triangle} Effective interaction
vertex induced by quantum gravity effects.}
\end{center}
\end{figure}
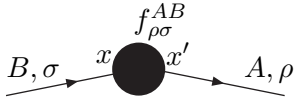

The possibility of neutrino oscillations arises, in general,  when
terms are added to the neutrino sector of the Standard Model
Lagrangian, such that the eigenstates of the total Hamiltonian
are not coincident with, but rather
linear combinations of the different neutrino flavors.
In the simplified case of two flavors, and a single
oscillation inducing interaction, a neutrino of flavor $\alpha$
can oscillate into a neutrino of flavor $\beta$ after having
traveled for a distance $X$, with a probability that is usually
expressed in terms of a mixing angle $\theta$ and an oscillation
length $L$
\begin{eqnarray}\label{prob}
P_{\alpha\rightarrow\beta}=\sin^2\left(2\theta\right)
\sin^2\left[\frac{\pi X}{L}\right]\, .
\end{eqnarray}
The oscillation length $L=\frac{2\pi}{E_A-E_B}$, is defined in terms of
the difference of the states' energy $E_A-E_B$ and does not
depend (for this case) on the mixing angle $\theta$. The mixing
angle determines only the amplitude of the oscillations.
In the case of more than two neutrino flavors, eq.~(\ref{prob}) becomes
more complicated, and includes more than one oscillatory term.
Since eq.~(\ref{prob}) contains the essential physical
ingredients of the oscillation phenomenon, and is much simpler
than the corresponding expression in the general case, we will
concentrate on the two flavor case.

If oscillations are induced by more than one interaction, $L$ should be
replaced by the total oscillation length $L_{tot}$, which is given by
a combination of the individual oscillation lengths $L_n$ that each
individual interaction would have induced on its own. In the two flavor case
($\theta_{n}$ are the mixing angles of each single effect) one
obtains \cite{ColGl2}
\begin{eqnarray}\label{Ltot}
L_{\rm tot}^{-1}=\frac{\sum\limits_{n}L_{n}^{-1} \cos 2\theta_{n}}
{\cos 2\theta_{\rm tot}}\, ,
\end{eqnarray}
from which it is clear that in general the total oscillation
length is dominated by the shortest one. This feature remains valid also
for the case of more than two flavors, though eq.~(\ref{Ltot})
is replaced by a more complicated expression.

Thus, if we are interested in a case in which one oscillation inducing
term (say $L_{j}$) dominates, we may replace $L$ by $L_{j}$
instead of using $L=L_{\rm tot}$  in eq.~(\ref{prob}),
and use the factorization property
of eq.(\ref{prob}) which implies that the
oscillation length does not depend on the mixing angles.

The most studied oscillation mechanism is due to the presence of
neutrino mass. For this case, LI implies that the oscillation
length for ultrarelativistic neutrinos is given by $ L\sim
\frac{4\pi E}{\Delta m^2}$, ${\Delta m^2}$ being the neutrino
squared mass difference and here $c=1=\hbar$. In fact, every LI
term in the effective action will give the same oscillation length
energy dependence as a mass term. On the other hand, energy
dependence different from $L\propto E$ is a smoking-gun signal of
violation of LI. Several examples of LIV effects have appeared in
the literature: $L\propto E^0$ \cite{DesGas,ColGl1}, $L\propto
E^{-1}$ \cite{Gasp,ColGl2} and also $L\propto E^{-2}$
\cite{eichler,AMU} and $L\propto E^{-3}$ \cite{Adunas:2000zm}.

Recent SuperKamiokande (SK) results indicate that a $\mu$-$\tau$
mass mixing mechanism is the best candidate for explaining the
observed $\mu$-$e$ anomaly in the atmospheric neutrino
flux\cite{SKK1}. The analysis of energy dependence of SK data
excludes LIV effects as the primary source for the observed
$\mu$-$e$ anomaly. Furthermore, by analyzing models in which LIV
terms are present in addition to the mass mixing, it is possible
to put stringent upper limits on their strength\cite{FLMS}. We
show below that existing experimental data are already sufficient
to provide important clues about the nature of neutrino
interaction at the Planck scale, and that with better analysis
and with more data LIV terms could be ruled out to high accuracy.
Our claim is based on a preliminary analysis of the $L\propto
E^{-2}$  case, which are induced only by nonrenormalizable terms
in the neutrino effective Lagrangian. Dimensional analysis shows
that renormalizable terms in the neutrino effective action induce
oscillation lengths proportional to $E^{-1}$ , or to a
non-negative power of E. In particular, all possible
renormalizable and rotation invariant mixing terms in the
neutrino effective action induce oscillation lengths proportional
either to $E^0$, to $E$, or to $E^{-1}$. This can be checked by a
case-by-case analysis of the dispersion relation induced by all
the operators that are bilinear in the neutrino field, and have
mass dimension not exceeding four. A complete set of such
operators is  $\bar{\psi}\gamma^0\psi$,
$\bar{\psi}\gamma^5\gamma^0\psi$, $\bar{\psi}\partial_t\psi$,
$\bar{\psi}\gamma^5\partial_t\psi$,
$\bar{\psi}\gamma^0\partial_t\psi$,
$\bar{\psi}\gamma^5\gamma^0\partial_t\psi$,
$\bar{\psi}\gamma^{0i}\partial_{i}\psi$,
$\bar{\psi}\gamma^{ij}\partial^{k}\epsilon_{ijk}\psi$. (We have
not included in this list operators that can be obtained as
linear combinations of them and LI terms). It follows that only
terms that are both nonrenormalizable and LIV can be the origin
of  $L\propto E^{-2}$.

We now wish to show that generic mixing angles that are induced
are large. Assuming that gravity is ``flavor blind" the
interaction Hamiltonian induced by the terms in
eq.(\ref{livint}), expressed in flavor basis is a ``democratic"
matrix $ H_{\rm int}=\left(
\begin{array}{cc}
  g(E) & g(E) \\
  g(E) & g(E)
\end{array}
\right)$, and the total Hamiltonian (including possibly mass
mixing) in the flavor basis could  be expressed as $
H_{\alpha\beta}=\left(
\begin{array}{cc}
  h_{11}+g(E) & h_{12}+g(E) \\
  h_{21}+g(E) & h_{22}+g(E)
\end{array}
\right), $ where $h_{12}=h_{21}$. In the limit that the energy
difference between the Hamiltonian eigenstates $\Delta
E=|E_1-E_2|$ is dominated by $g(E)$, the Hamiltonian eigenvectors
are given by $\frac{1}{\sqrt{2}}\left(
\begin{array}{c}
  1 \\
  1
\end{array}\right)$,
$\frac{1}{\sqrt{2}}\left(
\begin{array}{c}
  1 \\
  -1
\end{array}\right)$, and  therefore mixing is maximal. In
case $|g(E)|>|h_{12}|, |h_{11}-h_{22}|$ we therefore expect a
large mixing angle. Since the SK (and possibly also solar
neutrino) data appear to imply a large mixing angle due to LI
effects, at least among some pairs of neutrino types, we infer
that $h_{12} \ge |h_{11}-h_{22}|$, so that $|g(E)|>|h_{12}|$
implies $|g(E)|>|h_{11}-h_{22}|$ and is by itself sufficient to infer a
large mixing angle.

As a simple explicit example of the impact of a local term
resulting from the interaction Lagrangian (\ref{livint}),
consider adding a dimension five LIV interaction term to the
standard kinetic term of massless neutrinos,
\begin{eqnarray}\label{term}
{\cal L} =i\delta^{AB}\bar{\psi}_{A}\partial\hspace{-.08in}\slash
\ \psi_B -\frac{S^{AB}}{M}\bar{\psi}_{A}\partial_{t}\partial_{t}
\gamma^{0}\psi_{B},
\end{eqnarray}
with $S^{AB}=diag(S_A, S_B)$. To obtain $S^{AB}$ from the second
moments of $f^{AB}_{\rho\sigma}$ we have used the
$\gamma$-matrices as a basis to 4 by 4 matrices. In terms of the
``democratic" interaction Hamiltonian which we
have discussed previously one finds $g(E)=\frac{E^2}{2M}(S_{A}-S_{B})$.

As we have argued, an interaction term as in eq.(\ref{term}) is
generic to every nonrenormalizable theory that is also LIV. We
will explicitly show that the resulting oscillation length
satisfies $L\propto E^{-2}$. Other terms such as
$\frac{1}{M}\bar{\psi}\partial_i\partial_i\gamma^0\psi$,
$\frac{1}{M}\bar{\psi}\partial_t\partial_i\gamma^i\psi$,
$\frac{1}{M}\bar{\psi}\partial_t\partial_t\gamma^5\gamma^0\psi$,
$\frac{1}{M}\bar{\psi}\partial_i\partial_i\gamma^5\gamma^0\psi$,
$\frac{1}{M}\bar{\psi}\partial_t\partial_i\gamma^5\gamma^i\psi$
also produce $L\propto E^{-2}$. Recall that in this case the
oscillation length depends only on energy difference between the
oscillating states, and does not depend on the mixing angle
between flavor eigenstates which determines only the amplitude of
oscillations.

The equation of motion for the physical eigenstates  is
$\left(p_{\mu}\gamma^{\mu} -
\frac{S_{A,B}E^2}{M}\gamma^0\right)\psi_{A,B}=0\, ; $ by
multiplying this equation by $p_{\mu}\gamma^{\mu} -
\frac{S_{A,B}E^2}{M}\gamma^0$ and using the anticommutation
relations of gamma matrices, we obtain $ E^2 \left(1 -
\frac{S_{A,B} E}{M}\right)^2 - {\bf p}^2=0, $ and therefore, to
first order in $S_{A,B} E/M$ $E\sim|{\bf p}|\left(1 +
\frac{S_{A,B} E}{M}\right)$, from which it follows that
\begin{eqnarray}\label{L-2}
L\sim\frac{2\pi}{S_{A}-S_{B}}\frac{M}{E^2}\, ,
\end{eqnarray}
independently of the mixing angle between flavor eigenstates.

A LIV term could also arise in an otherwise LI theory, such as
critical string theory, via spontaneous breaking of Lorentz
symmetry. For example, a LIV contribution of the form (\ref{term})
could be generated, when a field (which could be the dilaton, or
any other of the moduli fields of string theory) is slowly
rolling in a background of a non-vanishing gravitational
potential. But in this case, the breaking would be proportional
to the time derivative of the field $(\dot{\Phi}/M)$ whose
magnitude is severely constrained. If $\dot{\Phi}$ is
non-vanishing then $\Phi$ has kinetic energy. The requirement
that the kinetic energy density of $\Phi$ is not larger than
universe closure density provides an incredibly stringent bound
$\dot{\Phi}/M_{\rm Planck}<10^{-61}$, which makes spontaneous
breaking effects completely undetectable. This conclusion extends
to all the terms that can produce an $L\propto E^{-2}$ behavior
since all of them must contain at least one time (or space)
derivative of a field, and therefore are subject to the same (or
similar) phenomenological constraints.

Because all nonrenormalizable theories are expected to generate
higher derivative terms at the cutoff scale, and because such
terms are not necessarily LI unless the physical cutoff mechanism
itself is intrinsically LI, as in critical string theory, a
detection (or exclusion) of $L\propto E^{-2}$ carries with it
information about the high energy/short-distance properties of
the theory. In contrast, the detection of an energy dependence
$L\propto E^{-1}$ produced by renormalizable terms  (though in
any case extremely interesting) would not carry with it any such
information.  If LIV at the cutoff scale is explicit instead of
spontaneous, typical mixing terms are expected to be of order
unity (and detectable, as we show below), simply because there is
no symmetry that protects them. It follows that the detection of
$L\propto E^{-2}$ effect would be a definite signal that
nonrenormalizable and LIV terms are present in the neutrino
effective action. We conclude that the detection of   $L\propto
E^{-2}$ dependence in neutrino oscillations would constitute
strong evidence against critical string theory, in which the
ultraviolet cutoff is realized in a LI way. Conversely, an
experimental exclusion of such an effect should be considered as
vindication of critical string theory as compared to other Planck
scale models that violate Lorentz symmetry.

If one evaluates eq.(\ref{L-2}), with $E \sim 1$ GeV, $M\sim
M_{\rm Planck}$, and $S_A-S_B=O(1)$ as expected in explicit LIV
where $g(M)\sim M$, one
obtains an oscillation length of about 10 km. Typical high end of
the energy range of solar neutrino observatories is about 10 MeV
which results in an oscillation length of about $10^{-3}$ AU.
These estimates suggests that if the mixing is close to maximal
mixing $\sin^2(2\theta)=1$, then previous neutrino oscillation
experiments CHORUS\cite{cho} and NOMAD\cite{nom}, and the
currently operating experiments SK\cite{SKK1}, K2K\cite{K2K}, and
SNO\cite{snofrstres}, are already able to detect the proposed
effect if $S_{A,B}$ is larger than about $10^{-2}$. In this
context, the existing data must be interpreted as a null result,
$S_{A,B}$ less than about $10^{-2}$, or very small mixing
$\sin^2(2\theta)\ll 1$ or both, accordingly to equation (\ref{prob}), as
the analysis of the energy dependence of the oscillation length of SK
data shows that the observed oscillation is due to mass mixing $L\propto
E$, rather than to LIV terms\cite{FLMS}. Planned experiments MINOS and
CNGS\cite{exps} will be able to strengthen and verify these
findings.  The situation is summarized in table I which assumes
maximal mixing. As can be seen from the table, we can already
conclude without further analysis that the actual upper bound on
$\alpha$ is about $10^{-3}$ for maximal mixing, and that the
forthcoming experiments are potentially capable of improving this
limit by at least one order of magnitude.
Translating this bound on $\alpha$ into a bound on $M$, we obtain
$M>10^3 M_{\rm Planck}$. It is worth noticing that the
experimental sensitivity for other possible sources of nonrenormalizable LIV,
such as  the detection of time delays of photons from distant astrophysical
sources, turns out to be significantly lower $M\sim M_{\rm Planck}$ (see
Table I in \cite{Ellis2}).
A more detailed
analysis could enhance the sensitivity of neutrino oscillation
experiments compared to the estimates in the table, as shown in
Fig.~2, and determine exclusion regions in
$\alpha,\sin^2(2\theta)$ plane.

\begin{table}[ht]
\begin{center}
\begin{tabular}{||l|l|l|l|l|l||}
\hline
\hline
EXP. & STATUS & $\langle E\rangle$ (GeV) & $L$ (Km) & $X$ (Km) & $\alpha$ \\
\hline
CHORUS & closed 1997 & 26 & $10^{-2}$  & $0.85$ & $10^{-2}$\\
\hline
NOMAD & closed 1999 & 24 & $10^{-2}$  & $0.94$ & $10^{-2}$\\
\hline \hline
SK & operating & 1.3 & 10 & 10-$10^4$ & 1-$10^{-3}$\\
\hline
K2K & operating & 1.3 & 10 & 250 & $10^{-2}$\\
\hline
SNO & operating & $0.008$ & $10^{5}$ & $ 10^{8}$ & $10^{-3}$\\
\hline \hline
MINOS & starting 2003 & 15 & 0.1 & 730 & $10^{-4}$\\
\hline
CNGS & starting 2005 & 17 & 0.1 & 732 & $10^{-4}$\\
\hline
\hline
\end{tabular}
\vspace{.2in} \caption{ \label{tab1}  Shown for each experiment
are its operation status, mean value of observed neutrino energy,
oscillation length according to (\ref{L-2}) (with $S_A-S_B =1$),
typical neutrino flight distance $X$, and the ratio $\alpha=L/X$.
Parameter $\alpha$ can be thought of as the value of $S_A-S_B$ for
which $L=X$ for each experiment, that is, the lowest value of
$S_A-S_B$ to which each experiment is sensitive, assuming maximal
mixing.}
\end{center}
\end{table}

Furthermore, all these experiments observe (or plan to observe) a
broad spectrum of incoming neutrino energies, from about 10 MeV to
few hundreds GeV or more, so in fact each experiment sets a
stronger constraint on nonrenormalizable LIV terms as the high
energy end of its range is exploited. In the case of
SuperKamiokande the flight-distance also varies - from a few tens
of Kilometers to about $10^4$ Km. Thus, better analysis of
available data, and additional data from planned experiments will
allow the significant strengthening of the bound on $S_{A,B}$,
perhaps down to $10^{-8}$.
\begin{figure}[ht]
\centerline{\psfig{figure=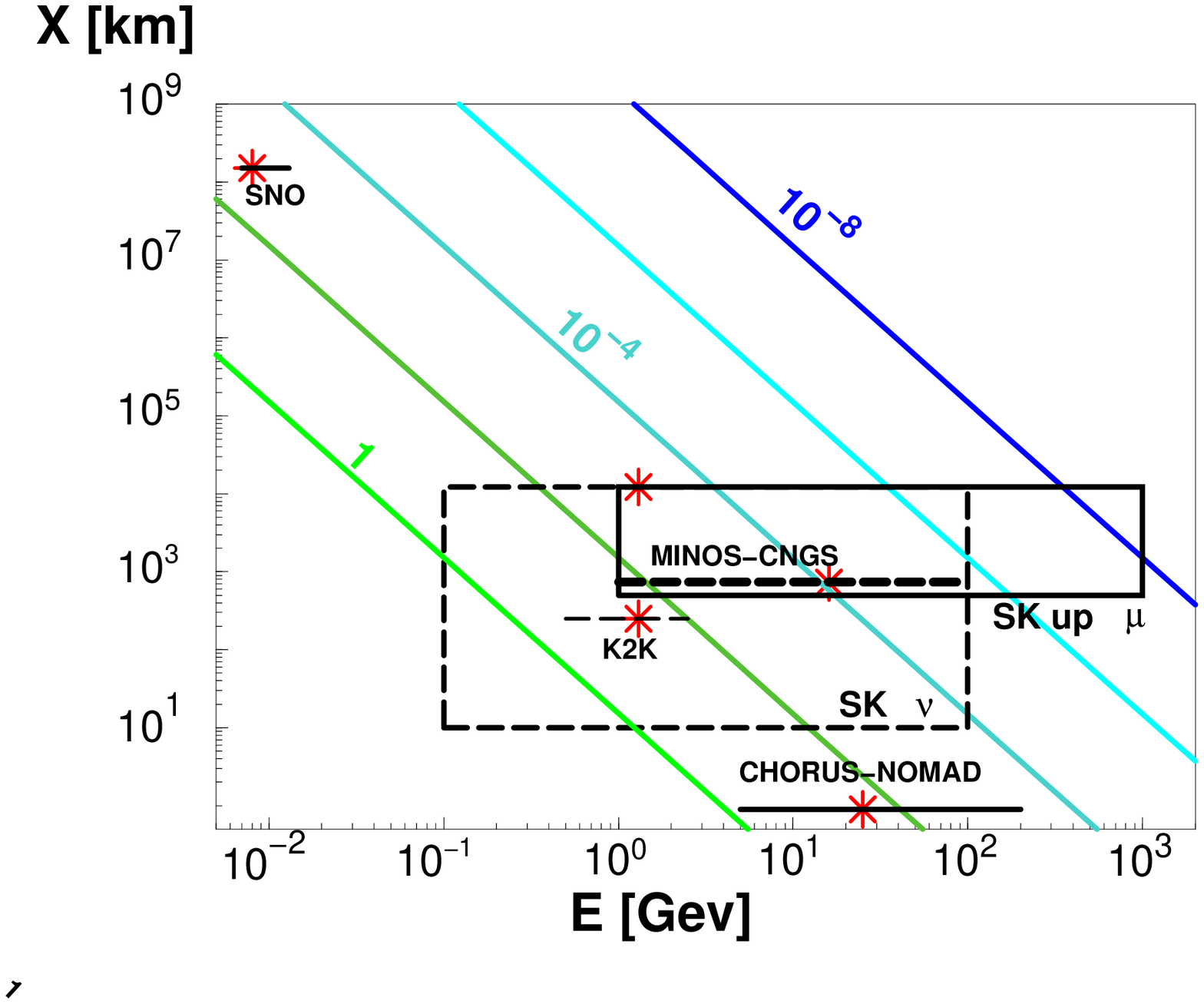, width=3in, angle=0}}
\caption{ \label{fig1} Shown are neutrino energy ranges and flight
distances for each experiment; the diagonal constant $\alpha$
lines indicate expected sensitivity. The SK reach is represented
by the two rectangles: the dashed one outlines the energy range
and flight distance of detected $\mu$ neutrinos, while the solid
one is for up-going muons.  The stars display the data reported
in table I, and mark neutrino energies for which luminosity is
maximal.}
\end{figure}
Figure~2 demonstrates that the limits reported in table I can be
improved by the analysis of the high-energy part of the neutrino
spectra. This is in contrast to the mass mixing case for which L
is increasing with energy, where the best bounds are obtained by
the lowest energy tail of neutrino spectra. Up-going muons in the
SuperKamiokande experiments, reaching energies of $10^3$ GeV ,
can provide the best bound on $\alpha$ at about $10^{-8}$, while a
value of $10^{-6}$  can potentially be reached by MINOS and CNGS.
A more detailed analysis is required to determine the highest
energy at which enough data can be accumulated.

We conclude that existing bounds are at most barely compatible
with the existence of a LIV ultraviolet cutoff at the Planck
scale. Strengthening of these bounds by better analysis and
additional data from future experiments would be an imminent
vindication of critical string theory, and a strike against
models that allow explicit breaking of the Lorentz symmetry at
the cutoff scale.


This research  was supported  by Arnow Chair of Astrophysics and by an
Adler Fellowship granted by the Israel Science Foundation (DE)
and by  Israel Science Foundation grant 174/00-2 (RB and SF).
SF is also supported by Della Riccia Foundation and the Kreitman Foundation.

\end{document}